\documentclass[reprint,twocolumn,aps,prl]{revtex4-1}
	\usepackage{graphicx}%
	\usepackage{dcolumn}%
	\usepackage{bm}%
	\usepackage{amsmath}
	\usepackage[colorlinks=true,linkcolor=blue]{hyperref}
\begin{document}
	\title{Towards Faster InP Photonic Crystal All-Optical-Gates}
	\author{Gregory Moille, Sylvain Combri\'e and Alfredo de Rossi}
	\affiliation{Thales Research and Technology, 1 Avenue A. Fresnel, 91767
Palaiseau, France}
	\author{Kerstin Fuchs, Matusala Yacob and Johann Peter Reithmaier}
	\affiliation{Institute of Nanostructure Technologies and Analytics, CINSaT, University of Kassel 34132 Kassel, Germany}
	\date{\today}%

	\begin{abstract}
	 	We demonstrated a two-fold acceleration of the fast time constant characterising the recovery of a P-doped Indium-Phosphide Photonic Crystal all-optical gate. Time-resolved spectral analysis is compared with a three-dimensional drift-diffusion model for the carrier dynamics, demonstrating the transition from the ambipolar to the faster minority carrier dominated diffusion regime. This open the perspective for faster yet efficient nanophotonic all-optical gates.
	\end{abstract}

\maketitle
\section{Introduction}

	\noindent The original motivation for all-optical signal processing is the intrinsic very high speed of the nonlinear optical processes. The ultra-fast power-dependent phase accumulation in a single-mode optical fibre can be converted into amplitude-modulation using a suitable interferometric configuration, such as the nonlinear optical loop mirror\cite{Doran1988}, which had indeed been used extensively for ultra-fast all-optical signal processing. More recently, the emphasis shifted into decreasing the overall power consumption, which is critical in optical interconnects. An important result, in this context, is the drastic decrease of energy required for all-optical switching in photonic crystal resonators, owing to the strong optical confinement\cite{Nozaki2010}. 
	Crucial here is also the very strong nonlinear effect resulting from the excitation of free-carriers and the consequent change of the refractive index\cite{Bennett1990}. This nonlinear response substantially differs from the Kerr effect since it is non-instantaneous. Carriers accumulate (and the refractive index changes) as the pump is being absorbed and then relax with a time which is related to the diffusion and the surface recombination. It has been shown that the strong gradient in the distribution of the photo-excited carriers in Indium Phosphide and in Silicon PhC cavities is responsible for a very fast diffusion, which amounts to a response which is much faster\cite{Tanabe2008,Nozaki2010} than the typical carrier lifetime in these materials. Thus, all-optical signal processing has been demonstrated  InP based PhC cavities, namely all-optical sampling\cite{Combrie2013}, wavelength conversion\cite{yu_nonlinear_2014,bazin_ultrafast_2014} and non-reciprocal transmission\cite{yu_nonreciprocal_2015} at a repetition rate of 10 GHz.\\
	It is however desirable to improve speed further, which necessarily implies to modify the material properties. It is well known that recombination rate can be improved by increasing the number of trap states. This has also been demonstrated with Silicon photonic crystal\cite{Tanabe2010} and micro-ring cavities\cite{Waldow2008}. Another possibility is to trap carriers at the surface using a quantum well, where recombination is much more effective\cite{bazin_ultrafast_2014}. Here we investigate the possibility of accelerating the diffusion process itself.\\
	PhC gates in Silicon and Indium Phosphide \cite{Tanabe2008,Yu2013,Nozaki2010} owe their fast response to a diffusion process which depends on the ambipolar diffusion constant $D_a = (D_n^{-1}+D_p^{-1})^{-1}$. It is apparent that, in this respect, Silicon and InP are expected to behave similarly, because $D_a \approx 5 cm^2/s $ is dominated by the slow holes in InP. In contrast, when InP is doped with acceptors, diffusion is regulated by minority carriers (electrons) with $D_n \approx 130 cm^2/s$, way larger, thus diffusion is expected to be much faster\cite{Sze1981}.\\ 
	Based on this, we have fabricated a nonlinear photonic crystal gate based on P-doped InP.  Acceptors, with density $N_a=5\times10^{16}\mathrm{cm}^{-3}$ are created \textit{in situ} during the Molecular Beam Epitaxy. This doping level is to be compared to the typical injection level in these structures.
	

\section{Sample design and fabrication}
		\begin{figure}[!h]
			\begin{center}
				\includegraphics[width=\columnwidth]{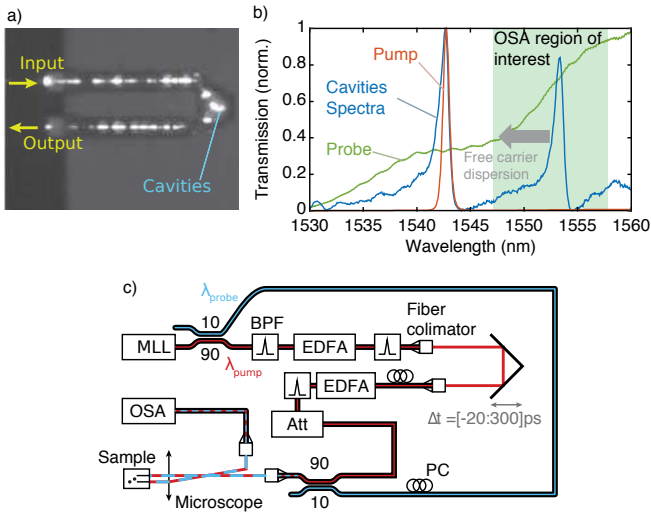}
				\caption{\label{fig:Sample} a) Infrared image of the sample from the top. The propagation of the light to and from the resonators is shown when the wavelength is set at the resonance. b) Transmission, pump and probe spectra.  c) Experimental set-up. In red the pump path which goes through the delay line, in blue the probe path.}
			\end{center}
		\end{figure}	

	\noindent The design of the device consists in a dual-mode resonator obtained by coupling two $H_0$ photonic crystal cavities\cite{zhang_small-volume_2004} patterned  on a 250 nm thick InP suspended slab. The period of the photonic crystal is $a=465$ nm, and the four nearest holes are displaced along  orthogonal directions by $d_x=0.16 a$  ($\mathbf{K}$ direction) and $d_y=0.07 a$.  The two cavities are spaced by the vector $\mathbf{R}=3\mathbf{a}_1+3\mathbf{a}_2$ ($\mathbf{M}$ direction). This design is known as a \textit{Photonic Molecule} and was introduced in InP all-optical gates recently\cite{Combrie2013}.

	The main particularity here is that the resonator itself is made very close to the input and output free space couplers, which are located on the same side and therefore accessed with the same objective lens (fig.~\ref{fig:Sample}a,c). As a result, the transmission spectra do no show any spectral features related to disorder-induced scattering, thereby limiting distortion and insertion losses.
	The device is fabricated by e-beam patterning a resist used to define a silica hard mask. The pattern is then transferred to the III-V semiconductor by Inductively Coupled Plasma (ICP) etching. The sacrificial InGaAsP underlying layer is removed by chemical etching.\\
	The fundamental resonances of the $H_0$ cavities are split by about 10 nm, which allows a comfortable spectral separation between the  pump and the probe.\\

	\section{Time-resolved spectral analysis}
	
	The recovery dynamics of the all-optical gate, specifically, the evolution of instantaneous detuning, is measured with a time-resolved spectral technique, consisting in recording the transmission (or the reflection) spectra of a short probe pulse as \textit{withe light}, as a function of its delay relative to the pump. This is similar to\cite{bristow_ultrafast_2003,raineri_ultrafast_2004}, except that here pump and probe are both obtained by a femto-second fibre laser (Optisiv) and, consequently, lie in the Telecom spectral domain, as in ref.\cite{Combrie2013}. Here, InP is transparent and the residual absorption can safely be neglected. Free carriers are generated by two-photon  absorption in the cavity only. The probe is approximately 80 nm broad and almost flat, and the duration is approximately 2 ps (Fig. \ref{fig:Sample}b). The pump is obtained after filtering and amplification  of the femtosecond laser output and the relative delay is controlled mechanically with a translation stage (Fig. \ref{fig:Sample}). The spectral power density of the probe is at least 2 orders of magnitude below the pump. Moreover, the same measurement was repeated by varying the probe level to confirm the absence of any influence on the dynamics. The output signal is send to and Optical Signal Analyser (OSA) and the region of interest is restrained around $\lambda_{probe}$(Fig. \ref{fig:Sample}c).
	\begin{figure}[!t]
		\begin{center}
			\includegraphics[width=1\columnwidth]{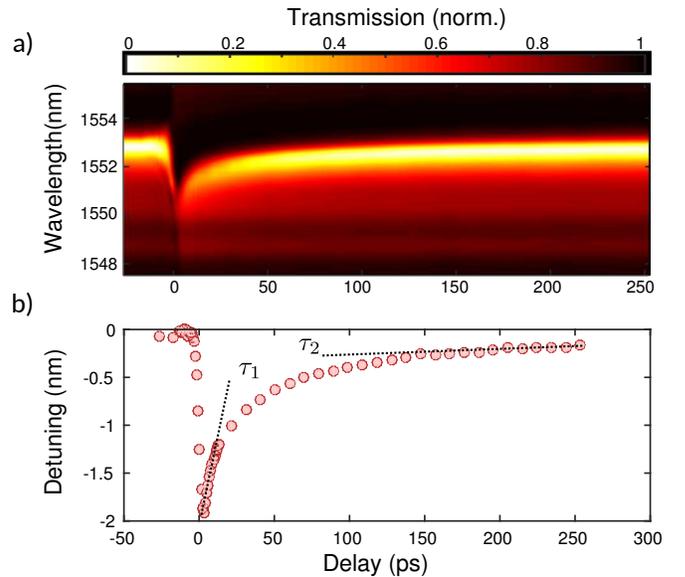}
			\caption{\label{fig:Map_lc} (a) Time-resolved experimental transmission spectra centered around the low-energy resonance (b) Extracted time-resolved instantaneous detuning and related slow and fast time constants $\tau_1$ and $\tau_2$}
		\end{center}
	\end{figure}
	The instantaneous cavity detuning is directly related to the carrier effective density, assuming that free-carrier induced index change is dominant, as is the case here. $\Delta \lambda \propto \frac{\partial N }{\partial n} N_{eff}$, where $N_{eff}\propto\int (n+p) W(\mathbf{r}) \mathrm{d}V$, i.e. the free carriers distribution weighted by the distribution of the electric energy $W(\mathbf{r})$ of the cavity mode (normalized).\\
	The time-resolved transmission spectra are represented as an intensity map (Fig.~\ref{fig:Map_lc}a), which clearly reveals the spectral blue shift of the resonance after the excitation by a 3 ps long pump pulse. An estimate of the instantaneous detuning  is obtained by calculating the first order moment of the spectral power density: $\lambda_c=\int \lambda S(\lambda)\mathrm{d}\lambda $, which is represented in Fig.~\ref{fig:Map_lc}b. Two time constants $\tau_1$ and $\tau_2$ are apparent.  The slow decay rate $\tau_2$ is related to the surface recombination, while the fast time constant $\tau_1$ corresponds to the fast diffusion dynamics\cite{Tanabe2008,heuck_heterodyne_2013,Nozaki2010}. 
%

\section{Results and discussion}
		\begin{figure}[!h]
			\begin{center}
				\includegraphics[width=0.8\columnwidth]{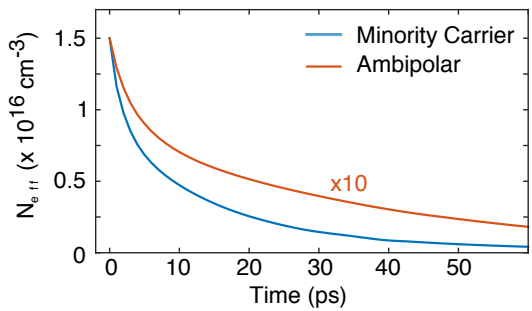}
				\caption{\label{fig:Simu_Neff} Calculated decay of the effective carrier density $N_{eff}$ in two different regimes. Low injection (blue), with $N_{eff}= 1.46 \times 10^{16}\mathrm{cm}^{-3}$ (blue) and high injection $N_{eff}= 1.46 \times 10^{17}\mathrm{cm}^{-3}$ (red, scaled down by a factor 10).}
			\end{center}
		\end{figure}	
	\noindent The carrier relaxation is either in the ambipolar regime or it is dominated by minority carriers (hence faster),  depending on whether the injected density is larger or smaller than the doping level. In order to appreciate this, we have developed a three-dimension finite element model of the drift-diffusion equations based on the COMSOL environment:
	\begin{align*}
		&\frac{\partial n}{\partial t} + \nabla \left(-D_n\nabla n + \mu_n n \nabla V \right)  = 0   \\
		&\frac{\partial p}{\partial t} + \nabla \left(-D_p\nabla p - \mu_p p \nabla V \right)  = 0  \\
		&(p-n-N_a)  = -\epsilon_r \frac{\epsilon_0}{q}\nabla^2 V
		\label{eq:dd}
	\end{align*}
	
	\noindent Here $n$ and $p$ represent the density of electrons and holes, $V$ is the electrical potential, $N_a=5\times10^{16}\mathrm{cm}^{-3}$ is the acceptor doping concentration and $D_n=130cm^2/s$ and $D_p=5cm^2/s$ and $\mu_n = 5400 cm^2/s$ and $\mu_n = 200 cm^2/s$. The initial condition is the carrier distribution, corresponding to the spatial distribution of the generation rate, which is proportional to the two-photon absorption rate, hence to $W(\mathbf{r})^2$. The result is shown in Fig. \ref{fig:Simu_Neff} and clearly reveals a strong change in the fast time constant (the slope of the relaxation) as the injection level is changed.\\

		\begin{figure}[!t]
			\begin{center}
				\includegraphics[width=0.9\columnwidth]{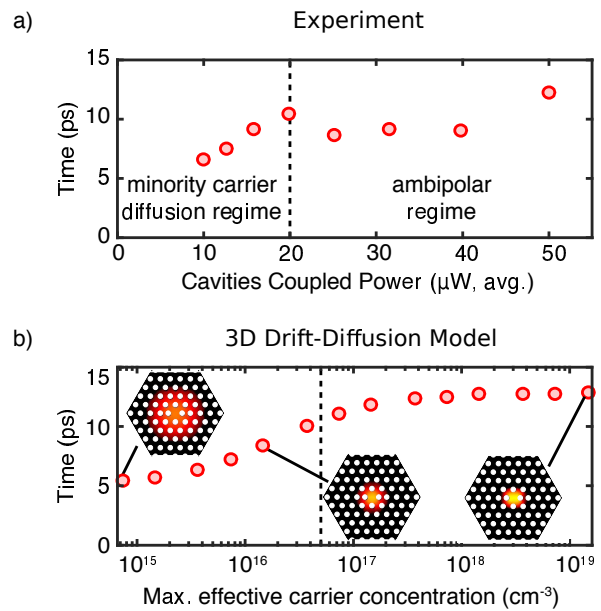}
				\caption{\label{fig:Fast_tau} (a) Measured fast time constant as a function of the input pump power coupled to the cavities. (b) Calculated time constant versus the maximum of the effective carrier density. The dotted line represents the level of p-doping. The distribution of the electrons at $t=10\mathrm{ps}$ is represented in the insets for different excitation levels.}
			\end{center}
		\end{figure}

	We now plot the measured fast time constant as a function of the coupled pump power (Fig. \ref{fig:Fast_tau}a), which is estimated by considering that linear transmission through the gate is about $80\%$ and that, by symmetry, the coupling coefficient is obtained by the square root of the transmission at resonance (about 2\%).\\
	Two regimes are clearly distinguishable: when the average input power is above 20 $\mu W$, which translates into a coupled pump energy of $80 fJ$, then $\tau_1$ is roughly constant and equal to 10 ps. Conversely, when the pump power is decreased below that level, then $\tau_1$ decreases steadily to about 6 ps for the weakest power injection ($40fJ$), only limited by measurement noise. This behavior is well reproduced by the our drift-diffusion model, which confirms that the cross-over is indeed related to when the injection level goes below the doping density, hence $5\times10^{16}cm^{-3}$. This is also apparent when considering the spatial distributions (Fig. \ref{fig:Fast_tau}b) of the electrons calculated after 10 ps, a time scale which is relevant for switching. In the minority carrier diffusion regime, electrons spread much farther away, which translates into a faster decrease of the effective carrier concentration (i.e. in the center of the cavity).

\section{Conclusion} 
	\noindent We demonstrate a two-fold acceleration of the fast time constant of the relaxation of the response of a p-doped InP all-optical gate based on a PhC resonator. The acceleration corresponds to the transition from ambipolar to minority carrier limited diffusion regime, which is confirmed by a 3D drift-diffusion model. The shortest experimental value of the fast time constant is 6 ps, half of the one in the ambipolar case (that is, in undoped InP structures). We point out that the doping level can be adjusted to set the transition to the minority carrier regime to a higher injection level, depending on the desired operation point. This opens new insight for ultra fast and energy efficient all-optical signal processing, such as time-domain demultiplexing, wavelength conversion and all-optical sampling of wide bandwidth microwave signals.

\hfill
\bibliographystyle{apsrev4-1}
\bibliography{Biblio}

\begin{thebibliography}{16}%
\makeatletter
\providecommand \@ifxundefined [1]{%
 \@ifx{#1\undefined}
}%
\providecommand \@ifnum [1]{%
 \ifnum #1\expandafter \@firstoftwo
 \else \expandafter \@secondoftwo
 \fi
}%
\providecommand \@ifx [1]{%
 \ifx #1\expandafter \@firstoftwo
 \else \expandafter \@secondoftwo
 \fi
}%
\providecommand \natexlab [1]{#1}%
\providecommand \enquote  [1]{``#1''}%
\providecommand \bibnamefont  [1]{#1}%
\providecommand \bibfnamefont [1]{#1}%
\providecommand \citenamefont [1]{#1}%
\providecommand \href@noop [0]{\@secondoftwo}%
\providecommand \href [0]{\begingroup \@sanitize@url \@href}%
\providecommand \@href[1]{\@@startlink{#1}\@@href}%
\providecommand \@@href[1]{\endgroup#1\@@endlink}%
\providecommand \@sanitize@url [0]{\catcode `\\12\catcode `\$12\catcode
  `\&12\catcode `\#12\catcode `\^12\catcode `\_12\catcode `\%12\relax}%
\providecommand \@@startlink[1]{}%
\providecommand \@@endlink[0]{}%
\providecommand \url  [0]{\begingroup\@sanitize@url \@url }%
\providecommand \@url [1]{\endgroup\@href {#1}{\urlprefix }}%
\providecommand \urlprefix  [0]{URL }%
\providecommand \Eprint [0]{\href }%
\providecommand \doibase [0]{http://dx.doi.org/}%
\providecommand \selectlanguage [0]{\@gobble}%
\providecommand \bibinfo  [0]{\@secondoftwo}%
\providecommand \bibfield  [0]{\@secondoftwo}%
\providecommand \translation [1]{[#1]}%
\providecommand \BibitemOpen [0]{}%
\providecommand \bibitemStop [0]{}%
\providecommand \bibitemNoStop [0]{.\EOS\space}%
\providecommand \EOS [0]{\spacefactor3000\relax}%
\providecommand \BibitemShut  [1]{\csname bibitem#1\endcsname}%
\let\auto@bib@innerbib\@empty
\bibitem [{\citenamefont {Doran}\ and\ \citenamefont {Wood}(1988)}]{Doran1988}%
  \BibitemOpen
  \bibfield  {author} {\bibinfo {author} {\bibfnamefont {N.~J.}\ \bibnamefont
  {Doran}}\ and\ \bibinfo {author} {\bibfnamefont {D.}~\bibnamefont {Wood}},\
  }\href {\doibase 10.1364/OL.13.000056} {\bibfield  {journal} {\bibinfo
  {journal} {Opt. Lett.}\ }\textbf {\bibinfo {volume} {13}},\ \bibinfo {pages}
  {56} (\bibinfo {year} {1988})}\BibitemShut {NoStop}%
\bibitem [{\citenamefont {Nozaki}\ \emph {et~al.}(2010)\citenamefont {Nozaki},
  \citenamefont {Tanabe}, \citenamefont {Shinya}, \citenamefont {Matsuo},
  \citenamefont {Sato}, \citenamefont {Taniyama},\ and\ \citenamefont
  {Notomi}}]{Nozaki2010}%
  \BibitemOpen
  \bibfield  {author} {\bibinfo {author} {\bibfnamefont {K.}~\bibnamefont
  {Nozaki}}, \bibinfo {author} {\bibfnamefont {T.}~\bibnamefont {Tanabe}},
  \bibinfo {author} {\bibfnamefont {A.}~\bibnamefont {Shinya}}, \bibinfo
  {author} {\bibfnamefont {S.}~\bibnamefont {Matsuo}}, \bibinfo {author}
  {\bibfnamefont {T.}~\bibnamefont {Sato}}, \bibinfo {author} {\bibfnamefont
  {H.}~\bibnamefont {Taniyama}}, \ and\ \bibinfo {author} {\bibfnamefont
  {M.}~\bibnamefont {Notomi}},\ }\href {\doibase 10.1038/nphoton.2010.89}
  {\bibfield  {journal} {\bibinfo  {journal} {Nat. Photonics}\ }\textbf
  {\bibinfo {volume} {4}},\ \bibinfo {pages} {477} (\bibinfo {year}
  {2010})}\BibitemShut {NoStop}%
\bibitem [{\citenamefont {Bennett}\ \emph {et~al.}(1990)\citenamefont
  {Bennett}, \citenamefont {Soref},\ and\ \citenamefont {Alamo}}]{Bennett1990}%
  \BibitemOpen
  \bibfield  {author} {\bibinfo {author} {\bibfnamefont {B.~R.}\ \bibnamefont
  {Bennett}}, \bibinfo {author} {\bibfnamefont {R.~A.}\ \bibnamefont {Soref}},
  \ and\ \bibinfo {author} {\bibfnamefont {J.~A. D. E.~L.}\ \bibnamefont
  {Alamo}},\ }\href {\doibase 10.1109/3.44924} {\bibfield  {journal} {\bibinfo
  {journal} {{IEEE} J. Quantum Electron.}\ }\textbf {\bibinfo {volume} {26}}
  (\bibinfo {year} {1990}),\ 10.1109/3.44924}\BibitemShut {NoStop}%
\bibitem [{\citenamefont {Tanabe}\ \emph {et~al.}(2008)\citenamefont {Tanabe},
  \citenamefont {Taniyama},\ and\ \citenamefont {Notomi}}]{Tanabe2008}%
  \BibitemOpen
  \bibfield  {author} {\bibinfo {author} {\bibfnamefont {T.}~\bibnamefont
  {Tanabe}}, \bibinfo {author} {\bibfnamefont {H.}~\bibnamefont {Taniyama}}, \
  and\ \bibinfo {author} {\bibfnamefont {M.}~\bibnamefont {Notomi}},\ }\href
  {\doibase 10.1109/JLT.2008.923638} {\bibfield  {journal} {\bibinfo  {journal}
  {J. Light. Technol.}\ }\textbf {\bibinfo {volume} {26}},\ \bibinfo {pages}
  {1396} (\bibinfo {year} {2008})}\BibitemShut {NoStop}%
\bibitem [{\citenamefont {Combri{\'e}}\ \emph {et~al.}(2013)\citenamefont
  {Combri{\'e}}, \citenamefont {Lehoucq}, \citenamefont {Junay}, \citenamefont
  {Malaguti}, \citenamefont {Bellanca}, \citenamefont {Reithmaier},
  \citenamefont {Bellanca}, \citenamefont {Trillo},\ and\ \citenamefont
  {Menager}}]{Combrie2013}%
  \BibitemOpen
  \bibfield  {author} {\bibinfo {author} {\bibfnamefont {S.}~\bibnamefont
  {Combri{\'e}}}, \bibinfo {author} {\bibfnamefont {G.}~\bibnamefont
  {Lehoucq}}, \bibinfo {author} {\bibfnamefont {A.}~\bibnamefont {Junay}},
  \bibinfo {author} {\bibfnamefont {S.}~\bibnamefont {Malaguti}}, \bibinfo
  {author} {\bibfnamefont {G.}~\bibnamefont {Bellanca}}, \bibinfo {author}
  {\bibfnamefont {J.~P.}\ \bibnamefont {Reithmaier}}, \bibinfo {author}
  {\bibfnamefont {G.}~\bibnamefont {Bellanca}}, \bibinfo {author}
  {\bibfnamefont {S.}~\bibnamefont {Trillo}}, \ and\ \bibinfo {author}
  {\bibfnamefont {L.}~\bibnamefont {Menager}},\ }\href {\doibase
  10.1063/1.4829556} {\bibfield  {journal} {\bibinfo  {journal} {Appl. Phys.
  Lett.}\ }\textbf {\bibinfo {volume} {103}},\ \bibinfo {pages} {193510}
  (\bibinfo {year} {2013})}\BibitemShut {NoStop}%
\bibitem [{\citenamefont {Yu}\ \emph {et~al.}(2014)\citenamefont {Yu},
  \citenamefont {Palushani}, \citenamefont {Heuck}, \citenamefont {Vukovic},
  \citenamefont {Peucheret}, \citenamefont {Yvind},\ and\ \citenamefont
  {Mork}}]{yu_nonlinear_2014}%
  \BibitemOpen
  \bibfield  {author} {\bibinfo {author} {\bibfnamefont {Y.}~\bibnamefont
  {Yu}}, \bibinfo {author} {\bibfnamefont {E.}~\bibnamefont {Palushani}},
  \bibinfo {author} {\bibfnamefont {M.}~\bibnamefont {Heuck}}, \bibinfo
  {author} {\bibfnamefont {D.}~\bibnamefont {Vukovic}}, \bibinfo {author}
  {\bibfnamefont {C.}~\bibnamefont {Peucheret}}, \bibinfo {author}
  {\bibfnamefont {K.}~\bibnamefont {Yvind}}, \ and\ \bibinfo {author}
  {\bibfnamefont {J.}~\bibnamefont {Mork}},\ }\href {\doibase
  10.1063/1.4893984} {\bibfield  {journal} {\bibinfo  {journal} {Applied
  Physics Letters}\ }\textbf {\bibinfo {volume} {105}},\ \bibinfo {pages}
  {071112} (\bibinfo {year} {2014})}\BibitemShut {NoStop}%
\bibitem [{\citenamefont {Bazin}\ \emph {et~al.}(2014)\citenamefont {Bazin},
  \citenamefont {Lengl\'{e}}, \citenamefont {Gay}, \citenamefont {Monnier},
  \citenamefont {Bramerie}, \citenamefont {Braive}, \citenamefont {Beaudoin},
  \citenamefont {Sagnes}, \citenamefont {Raj},\ and\ \citenamefont
  {Raineri}}]{bazin_ultrafast_2014}%
  \BibitemOpen
  \bibfield  {author} {\bibinfo {author} {\bibfnamefont {A.}~\bibnamefont
  {Bazin}}, \bibinfo {author} {\bibfnamefont {K.}~\bibnamefont {Lengl\'{e}}},
  \bibinfo {author} {\bibfnamefont {M.}~\bibnamefont {Gay}}, \bibinfo {author}
  {\bibfnamefont {P.}~\bibnamefont {Monnier}}, \bibinfo {author} {\bibfnamefont
  {L.}~\bibnamefont {Bramerie}}, \bibinfo {author} {\bibfnamefont
  {R.}~\bibnamefont {Braive}}, \bibinfo {author} {\bibfnamefont
  {G.}~\bibnamefont {Beaudoin}}, \bibinfo {author} {\bibfnamefont
  {I.}~\bibnamefont {Sagnes}}, \bibinfo {author} {\bibfnamefont
  {R.}~\bibnamefont {Raj}}, \ and\ \bibinfo {author} {\bibfnamefont
  {F.}~\bibnamefont {Raineri}},\ }\href {\doibase 10.1063/1.4861121} {\bibfield
   {journal} {\bibinfo  {journal} {Applied Physics Letters}\ }\textbf {\bibinfo
  {volume} {104}},\ \bibinfo {pages} {011102} (\bibinfo {year}
  {2014})}\BibitemShut {NoStop}%
\bibitem [{\citenamefont {Yu}\ \emph {et~al.}(2015)\citenamefont {Yu},
  \citenamefont {Chen}, \citenamefont {Hu}, \citenamefont {Xue}, \citenamefont
  {Yvind},\ and\ \citenamefont {Mork}}]{yu_nonreciprocal_2015}%
  \BibitemOpen
  \bibfield  {author} {\bibinfo {author} {\bibfnamefont {Y.}~\bibnamefont
  {Yu}}, \bibinfo {author} {\bibfnamefont {Y.}~\bibnamefont {Chen}}, \bibinfo
  {author} {\bibfnamefont {H.}~\bibnamefont {Hu}}, \bibinfo {author}
  {\bibfnamefont {W.}~\bibnamefont {Xue}}, \bibinfo {author} {\bibfnamefont
  {K.}~\bibnamefont {Yvind}}, \ and\ \bibinfo {author} {\bibfnamefont
  {J.}~\bibnamefont {Mork}},\ }\href {\doibase 10.1002/lpor.201400207}
  {\bibfield  {journal} {\bibinfo  {journal} {Laser \& Photonics Reviews}\
  }\textbf {\bibinfo {volume} {9}},\ \bibinfo {pages} {241} (\bibinfo {year}
  {2015})}\BibitemShut {NoStop}%
\bibitem [{\citenamefont {Tanabe}\ \emph {et~al.}(2010)\citenamefont {Tanabe},
  \citenamefont {Nishiguchi}, \citenamefont {Shinya}, \citenamefont
  {Kuramochi}, \citenamefont {Inokawa}, \citenamefont {Notomi}, \citenamefont
  {Yamada}, \citenamefont {Tsuchizawa}, \citenamefont {Watanabe}, \citenamefont
  {Fukuda}, \citenamefont {Shinojima}, \citenamefont {Itabashi}, \citenamefont
  {Tanabe}, \citenamefont {Nishiguchi}, \citenamefont {Shinya},\ and\
  \citenamefont {Kuramochi}}]{Tanabe2010}%
  \BibitemOpen
  \bibfield  {author} {\bibinfo {author} {\bibfnamefont {T.}~\bibnamefont
  {Tanabe}}, \bibinfo {author} {\bibfnamefont {K.}~\bibnamefont {Nishiguchi}},
  \bibinfo {author} {\bibfnamefont {A.}~\bibnamefont {Shinya}}, \bibinfo
  {author} {\bibfnamefont {E.}~\bibnamefont {Kuramochi}}, \bibinfo {author}
  {\bibfnamefont {H.}~\bibnamefont {Inokawa}}, \bibinfo {author} {\bibfnamefont
  {M.}~\bibnamefont {Notomi}}, \bibinfo {author} {\bibfnamefont
  {K.}~\bibnamefont {Yamada}}, \bibinfo {author} {\bibfnamefont
  {T.}~\bibnamefont {Tsuchizawa}}, \bibinfo {author} {\bibfnamefont
  {T.}~\bibnamefont {Watanabe}}, \bibinfo {author} {\bibfnamefont
  {H.}~\bibnamefont {Fukuda}}, \bibinfo {author} {\bibfnamefont
  {H.}~\bibnamefont {Shinojima}}, \bibinfo {author} {\bibfnamefont
  {S.}~\bibnamefont {Itabashi}}, \bibinfo {author} {\bibfnamefont
  {T.}~\bibnamefont {Tanabe}}, \bibinfo {author} {\bibfnamefont
  {K.}~\bibnamefont {Nishiguchi}}, \bibinfo {author} {\bibfnamefont
  {A.}~\bibnamefont {Shinya}}, \ and\ \bibinfo {author} {\bibfnamefont
  {E.}~\bibnamefont {Kuramochi}},\ }\href {\doibase 10.1063/1.2431767}
  {\bibfield  {journal} {\bibinfo  {journal} {Appl Phys Lett}\ }\textbf
  {\bibinfo {volume} {031115}},\ \bibinfo {pages} {2005} (\bibinfo {year}
  {2010})}\BibitemShut {NoStop}%
\bibitem [{\citenamefont {Waldow}\ \emph {et~al.}(2008)\citenamefont {Waldow},
  \citenamefont {Pl{\"o}tzing}, \citenamefont {Gottheil}, \citenamefont
  {F{\"o}rst}, \citenamefont {Bolten}, \citenamefont {Wahlbrink},\ and\
  \citenamefont {Kurz}}]{Waldow2008}%
  \BibitemOpen
  \bibfield  {author} {\bibinfo {author} {\bibfnamefont {M.}~\bibnamefont
  {Waldow}}, \bibinfo {author} {\bibfnamefont {T.}~\bibnamefont
  {Pl{\"o}tzing}}, \bibinfo {author} {\bibfnamefont {M.}~\bibnamefont
  {Gottheil}}, \bibinfo {author} {\bibfnamefont {M.}~\bibnamefont {F{\"o}rst}},
  \bibinfo {author} {\bibfnamefont {J.}~\bibnamefont {Bolten}}, \bibinfo
  {author} {\bibfnamefont {T.}~\bibnamefont {Wahlbrink}}, \ and\ \bibinfo
  {author} {\bibfnamefont {H.}~\bibnamefont {Kurz}},\ }\href {\doibase
  10.1364/OE.16.007693} {\bibfield  {journal} {\bibinfo  {journal} {Opt
  Express}\ }\textbf {\bibinfo {volume} {16}},\ \bibinfo {pages} {2254}
  (\bibinfo {year} {2008})}\BibitemShut {NoStop}%
\bibitem [{\citenamefont {Yu}\ \emph {et~al.}(2013)\citenamefont {Yu},
  \citenamefont {Palushani}, \citenamefont {Heuck}, \citenamefont {Kuznetsova},
  \citenamefont {Kristensen}, \citenamefont {Ek}, \citenamefont {Peucheret},
  \citenamefont {Oxenl{\o}we}, \citenamefont {Rossi}, \citenamefont {Yvind},\
  and\ \citenamefont {M{\o}rk}}]{Yu2013}%
  \BibitemOpen
  \bibfield  {author} {\bibinfo {author} {\bibfnamefont {Y.}~\bibnamefont
  {Yu}}, \bibinfo {author} {\bibfnamefont {E.}~\bibnamefont {Palushani}},
  \bibinfo {author} {\bibfnamefont {M.}~\bibnamefont {Heuck}}, \bibinfo
  {author} {\bibfnamefont {N.}~\bibnamefont {Kuznetsova}}, \bibinfo {author}
  {\bibfnamefont {T.}~\bibnamefont {Kristensen}}, \bibinfo {author}
  {\bibfnamefont {S.}~\bibnamefont {Ek}}, \bibinfo {author} {\bibfnamefont
  {C.}~\bibnamefont {Peucheret}}, \bibinfo {author} {\bibfnamefont {L.~K.}\
  \bibnamefont {Oxenl{\o}we}}, \bibinfo {author} {\bibfnamefont {A.~D.}\
  \bibnamefont {Rossi}}, \bibinfo {author} {\bibfnamefont {K.}~\bibnamefont
  {Yvind}}, \ and\ \bibinfo {author} {\bibfnamefont {J.}~\bibnamefont
  {M{\o}rk}},\ }\href {\doibase 10.1364/OE.21.031047} {\bibfield  {journal}
  {\bibinfo  {journal} {Opt. Express}\ }\textbf {\bibinfo {volume} {90}},\
  \bibinfo {pages} {477} (\bibinfo {year} {2013})}\BibitemShut {NoStop}%
\bibitem [{\citenamefont {Sze}\ and\ \citenamefont {Ng}(1981)}]{Sze1981}%
  \BibitemOpen
  \bibfield  {author} {\bibinfo {author} {\bibfnamefont {S.~M.}\ \bibnamefont
  {Sze}}\ and\ \bibinfo {author} {\bibfnamefont {K.~K.}\ \bibnamefont {Ng}},\
  }\href@noop {} {\emph {\bibinfo {title} {Physics of Semiconductor
  Devices}}},\ \bibinfo {edition} {2006th}\ ed.\ (\bibinfo  {publisher} {John
  Wiley \& Sons},\ \bibinfo {year} {1981})\BibitemShut {NoStop}%
\bibitem [{\citenamefont {Zhang}\ and\ \citenamefont
  {Qiu}(2004)}]{zhang_small-volume_2004}%
  \BibitemOpen
  \bibfield  {author} {\bibinfo {author} {\bibfnamefont {Z.}~\bibnamefont
  {Zhang}}\ and\ \bibinfo {author} {\bibfnamefont {M.}~\bibnamefont {Qiu}},\
  }\href {\doibase 10.1364/OPEX.12.003988} {\bibfield  {journal} {\bibinfo
  {journal} {Optics Express}\ }\textbf {\bibinfo {volume} {12}},\ \bibinfo
  {pages} {3988} (\bibinfo {year} {2004})}\BibitemShut {NoStop}%
\bibitem [{\citenamefont {Bristow}\ \emph {et~al.}(2003)\citenamefont
  {Bristow}, \citenamefont {Wells}, \citenamefont {Fan}, \citenamefont {Fox},
  \citenamefont {Skolnick}, \citenamefont {Whittaker}, \citenamefont
  {Tahraoui}, \citenamefont {Krauss},\ and\ \citenamefont
  {Roberts}}]{bristow_ultrafast_2003}%
  \BibitemOpen
  \bibfield  {author} {\bibinfo {author} {\bibfnamefont {A.~D.}\ \bibnamefont
  {Bristow}}, \bibinfo {author} {\bibfnamefont {J.-P.~R.}\ \bibnamefont
  {Wells}}, \bibinfo {author} {\bibfnamefont {W.~H.}\ \bibnamefont {Fan}},
  \bibinfo {author} {\bibfnamefont {A.~M.}\ \bibnamefont {Fox}}, \bibinfo
  {author} {\bibfnamefont {M.~S.}\ \bibnamefont {Skolnick}}, \bibinfo {author}
  {\bibfnamefont {D.~M.}\ \bibnamefont {Whittaker}}, \bibinfo {author}
  {\bibfnamefont {A.}~\bibnamefont {Tahraoui}}, \bibinfo {author}
  {\bibfnamefont {T.~F.}\ \bibnamefont {Krauss}}, \ and\ \bibinfo {author}
  {\bibfnamefont {J.~S.}\ \bibnamefont {Roberts}},\ }\href {\doibase
  10.1063/1.1598647} {\bibfield  {journal} {\bibinfo  {journal} {Applied
  Physics Letters}\ }\textbf {\bibinfo {volume} {83}},\ \bibinfo {pages} {851}
  (\bibinfo {year} {2003})}\BibitemShut {NoStop}%
\bibitem [{\citenamefont {Raineri}\ \emph {et~al.}(2004)\citenamefont
  {Raineri}, \citenamefont {Cojocaru}, \citenamefont {Monnier}, \citenamefont
  {Levenson}, \citenamefont {Raj}, \citenamefont {Seassal}, \citenamefont
  {Letartre},\ and\ \citenamefont {Viktorovitch}}]{raineri_ultrafast_2004}%
  \BibitemOpen
  \bibfield  {author} {\bibinfo {author} {\bibfnamefont {F.}~\bibnamefont
  {Raineri}}, \bibinfo {author} {\bibfnamefont {C.}~\bibnamefont {Cojocaru}},
  \bibinfo {author} {\bibfnamefont {P.}~\bibnamefont {Monnier}}, \bibinfo
  {author} {\bibfnamefont {A.}~\bibnamefont {Levenson}}, \bibinfo {author}
  {\bibfnamefont {R.}~\bibnamefont {Raj}}, \bibinfo {author} {\bibfnamefont
  {C.}~\bibnamefont {Seassal}}, \bibinfo {author} {\bibfnamefont
  {X.}~\bibnamefont {Letartre}}, \ and\ \bibinfo {author} {\bibfnamefont
  {P.}~\bibnamefont {Viktorovitch}},\ }\href {\doibase 10.1063/1.1788884}
  {\bibfield  {journal} {\bibinfo  {journal} {Applied physics letters}\
  }\textbf {\bibinfo {volume} {85}},\ \bibinfo {pages} {1880} (\bibinfo {year}
  {2004})}\BibitemShut {NoStop}%
\bibitem [{\citenamefont {Heuck}\ \emph {et~al.}(2013)\citenamefont {Heuck},
  \citenamefont {Combri\'{e}}, \citenamefont {Lehoucq}, \citenamefont
  {Malaguti}, \citenamefont {Bellanca}, \citenamefont {Trillo}, \citenamefont
  {Kristensen}, \citenamefont {Mo̸rk}, \citenamefont {Reithmaier},\ and\
  \citenamefont {de~Rossi}}]{heuck_heterodyne_2013}%
  \BibitemOpen
  \bibfield  {author} {\bibinfo {author} {\bibfnamefont {M.}~\bibnamefont
  {Heuck}}, \bibinfo {author} {\bibfnamefont {S.}~\bibnamefont {Combri\'{e}}},
  \bibinfo {author} {\bibfnamefont {G.}~\bibnamefont {Lehoucq}}, \bibinfo
  {author} {\bibfnamefont {S.}~\bibnamefont {Malaguti}}, \bibinfo {author}
  {\bibfnamefont {G.}~\bibnamefont {Bellanca}}, \bibinfo {author}
  {\bibfnamefont {S.}~\bibnamefont {Trillo}}, \bibinfo {author} {\bibfnamefont
  {P.~T.}\ \bibnamefont {Kristensen}}, \bibinfo {author} {\bibfnamefont
  {J.}~\bibnamefont {Mo̸rk}}, \bibinfo {author} {\bibfnamefont {J.~P.}\
  \bibnamefont {Reithmaier}}, \ and\ \bibinfo {author} {\bibfnamefont
  {A.}~\bibnamefont {de~Rossi}},\ }\href {\doibase 10.1063/1.4828355}
  {\bibfield  {journal} {\bibinfo  {journal} {Applied Physics Letters}\
  }\textbf {\bibinfo {volume} {103}},\ \bibinfo {pages} {181120} (\bibinfo
  {year} {2013})}\BibitemShut {NoStop}%
\end{thebibliography}%

\end{document}